\documentclass[12pt,useAMS,useAASTEX,usenatbib]{aastex}
\usepackage{emulateapj5}
\makeatletter
\newenvironment{inlinetable}{%
\def\@captype{table}%
\noindent\begin{minipage}{0.98\linewidth}\begin{center}\footnotesize}
{\end{center}\end{minipage}}

\newenvironment{apjemufigure}{%
\def\@captype{figure}%
\noindent\begin{minipage}{0.999\linewidth}\begin{center}}
{\end{center}\end{minipage}}
\makeatother

\def\l{{\ell}}
\def\lm{{\l m}}
\def\summ{\sum_{m=-\ell}^{\ell}}
\def\suml{\sum_{\ell=0}^{\infty}}
\def\dl{\Delta \l}
\def\dm{\Delta m}
\def\alm{a_{\lm}}
\def\ylm{Y_{\lm}}
\def\dl{{\Delta \l}}

\def\healpix{H{\sc ealpix }}
\def\glesp{G{\sc lesp }}
\def\wmap{\hbox{\sl WMAP~}}

\def\dilc{{\rm dilc}}

\def\etal{et al.}

\newcommand{\nbi}{{Niels Bohr Institute, Blegdamsvej 17,
DK-2100 Copenhagen, Denmark}}
\newcommand{\nott}{{ School of Physics \& Astronomy, University of
Nottingham, University Park, Nottingham NG7 2RD, United Kingdom}}

\slugcomment{Submitted to the Astrophysical Journal Letter}

\begin{document}

\title{Testing the Gaussian Random Hypothesis with the Cosmic Microwave Background Temperature Anisotropies in the 3-year WMAP Data}

\author{
Lung-Yih Chiang\altaffilmark{1},
Pavel D. Naselsky\altaffilmark{1},
Peter Coles\altaffilmark{2}
}

\altaffiltext{1}{\nbi}
\altaffiltext{2}{\nott}

\email{chiang@nbi.dk, naselsky@nbi.dk, peter.coles@nottingham.ac.uk}

\begin{abstract}
We test the hypothesis that the temperature of the cosmic microwave
background is consistent with a Gaussian random field defined on the
celestial sphere, using de-biased internal linear combination (DILC)
map produced from the 3-year \wmap data. We test the phases for
spherical harmonic modes with $\l \le 10$ (which should be the
cleanest) for their uniformity, randomness, and correlation with
those of the foreground templates. The phases themselves are
consistent with a uniform distribution, but not for $\l \le 5$, and the
differences between phases are not consistent with uniformity. For
$\l=3$ and $\l=6$,  the phases of the CMB maps cross-correlate with
the foregrounds, suggestion the presence of residual contamination
in the DILC map even on these large scales. We also use a
one-dimensional Fourier representation to assemble $\alm$ into the
$\Delta T_\l(\varphi)$ for each $\l$ mode, and test the positions of
the resulting maxima and minima for consistency with uniformity
randomness on the unit circle. The results show significant
departures at the 0.5\% level, with the one-dimensional peaks being
concentrated around $\varphi=180^\circ$. This strongly
significant alignment with the Galactic meridian, together with the
cross-correlation of DILC phases with the foreground maps, strongly
suggests that even the lowest spherical harmonic modes in the map
are significantly contaminated with foreground radiation.
\end{abstract}


\section{Introduction}
Since the release of the 1-year Wilkinson Microwave Anisotropy Probe
(\wmap) data \citep{wmapresults,wmapfg,wmappara,wmapcl,wmapng},
great efforts have been made to search and detect various possible
forms of non-Gaussianity of the cosmic microwave background (CMB)
temperature fluctuations
\citep{wmaptacng,coleskuiper,park,eriksenmf,santanderng,copi,romanng,hansen,mukherjee,larson,mcewen,edingburgh,mnn}.

Three years of data are now available and methods of foreground
cleaning have also been improved, so the \wmap team have produced a
new `de-biased' version of their internal linear combination map
(henceforth the DILC), which is claimed to be suitable for analysis
over the full sky for spherical harmonic modes up to $\l \le 10$
\citep{wmap3ycos,wmap3ytem}. The statistics of these low multipole
modes provide valuable information with cosmological significance,
particularly concerning statistical isotropy (or lack of it),
demonstrated by the CMB.

The 3-year data analyzed by the \wmap team is claimed to be
Gaussian. Non-Gaussianity, if detected, could result from primordial
origin \citep{nginflation}, possible foreground residues left over
after cleaning \citep{ndv03,ndv04,faraday,4n}, and/or correlated
instrument
noise. One will have to be very cautious if any
non-Gaussianity is detected before attributing it to a primordial
origin. The concept of internal linear combination to obtain a
reasonably clean CMB signal is to tune a set of weighting
coefficients in order to minimize the variance of the foregrounds.
Note that the variance is minimized but not eliminated. The
possibility that foreground signals remain in the DILC map to a
significant extent is therefore something that should be carefully
tested.

In this paper we apply a series of stringent tests of the Gaussian
hypothesis based on the behaviour of the phases of the spherical
harmonic modes in the data. Since these phases are highly sensitive
to the morphology \citep{morph} of the temperature pattern,
comparison between the phases of the DILC and the derived
foregrounds should give a sensitive indication of the presence of
contamination. Phase information can also be used to diagnose
departures from statistical homogeneity over the celestial sphere.

Using spherical harmonic phases in statistical tests involves some
subtleties.  For one thing, they are not rotationally invariant. In
other words a different choice of $z$-axis leads to a different
assignment of phases for the spherical harmonic modes of the same
pattern. This can be dealt with in a number of ways
\citep{coleskuiper}, but in the present context we choose to fix our
coordinate frame as that which makes most sense given the probable
behaviour of the foregrounds. In all the following we assume a
Galactic coordinate system; all phase information is interpreted
relative to this preferred frame. In doing this we attempt to ensure
that the detection of non-uniformity or non-randomness in the phases
can be interpreted more simply in terms of Galactic foregrounds.
Issues such as the claimed north-south asymmetry \citep{eriksenasym}
and the alignment of multipoles \citep{toh,oliveiracosta,schwarz,evil}, which
seem to persist in the \wmap 3-year data, are also measured in
Galactic coordinates,  so we hope to shed some additional light on
these peculiarities.

Owing to the visual similarity of 1-year ILC and 3-year DILC maps,
we will subject both maps to our analysis.

\begin{apjemufigure}
\centering
\centerline{\includegraphics[width=1.0\linewidth]{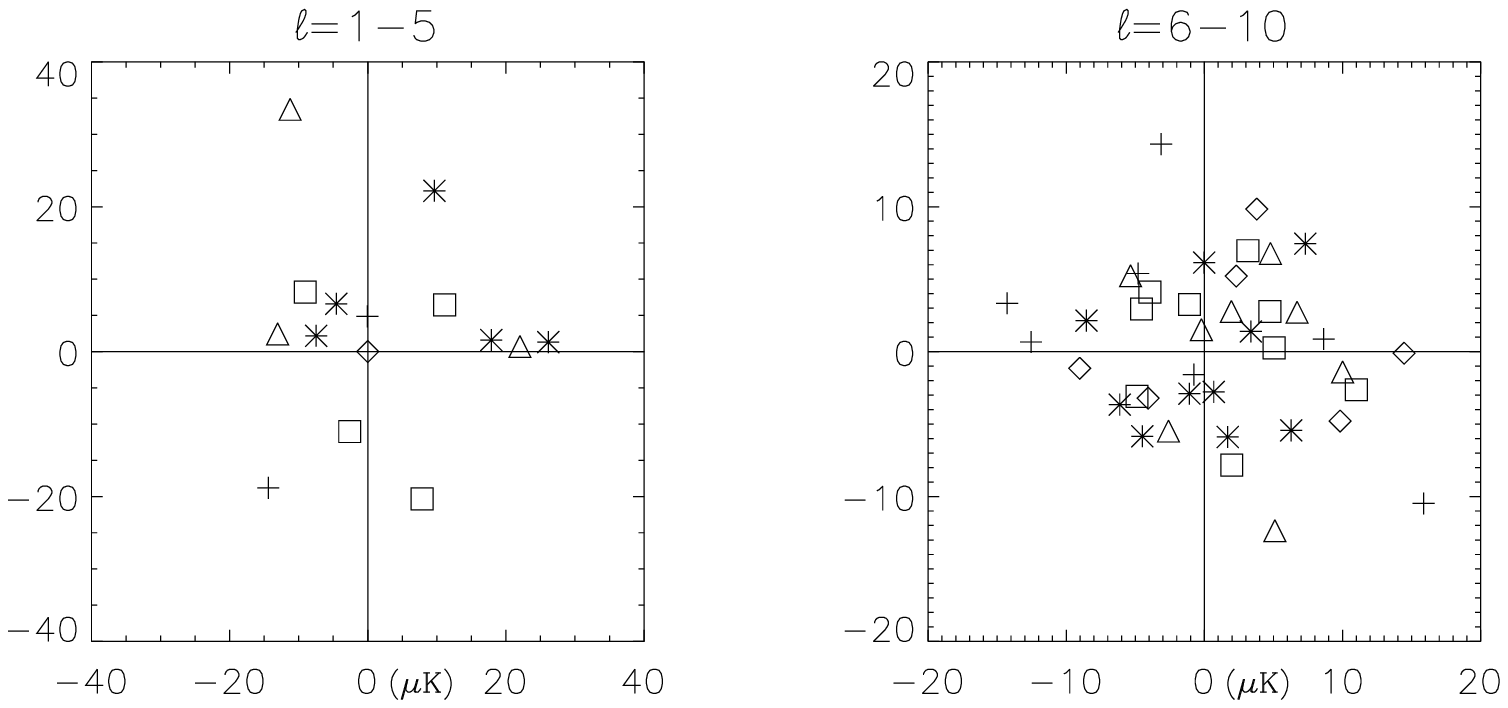}}
\caption{The $\alm$ for $\l \le 10$ are plotted on the Argand plane.
The amplitudes are in unit of $\mu$K and only $m>0$ modes are
displayed. The signs $\diamond$, $+$, $\triangle$, $\Box$ and $\ast$
represent $\l=1-5$ on the left, $\l=6-10$ on the right panel,
respectively. Note that 12 out of 15 phases of $\l\le5$ are in the
first two quadrants, among which six are clustered near 0 or $\pi$.}
\label{dist}
\end{apjemufigure}

\section{The Gaussian Random Hypothesis and the CMB}
The statistical characterization of CMB temperature anisotropies on
a sphere can be expressed as a sum over spherical harmonics:
\begin{equation}
\Delta T(\theta,\varphi)=\suml \summ \alm \ylm (\theta,\varphi),
\end{equation}
where the $\ylm(\theta,\varphi) $ are spherical harmonic functions,
defined in terms of the Legendre polynomials $P_\lm$ using
\begin{equation}
\ylm(\theta,\varphi)=(-1)^m \sqrt{\frac{(2\l+1)(\l-m)!}{4\pi(\l+m)!}}P_\lm(\cos\theta)\exp(i m \phi)
\end{equation}
and the $\alm$ are complex coefficients which can be expressed with
$\alm=|\alm| \exp(i \phi_{\lm})$. In standard cosmological models
(i.e. those involving the simplest forms of inflation) these
fluctuations constitute a realization of a statistically homogeneous
and isotropic Gaussian stochastic process, or random field, defined
over the celestial sphere \citep{bbks,be}. The formal definition of
such a Gaussian random field requires that the real and imaginary
parts of the $\alm$ are independent and identically distributed
according to a Gaussian probability density, so that the moduli
$|\alm|$ have a Rayleigh distribution and the phases $\phi_{\lm}$
are uniformly random on the interval $[0,2\pi]$. The Central Limit
Theorem virtually guarantees that the superposition of a large
number of harmonic modes will tend to a Gaussian as long as the
phases are random, and this furnishes a weaker definition of
Gaussianity. Moreover, statistical isotropy in general manifests
itself in phase properties. Because of the importance of phases in
both these definitions, we focus on their measured properties as
probes of departures from Gaussianity. Many methods have been
proposed to test the random phase hypothesis
\citep{phasemapping,meanchisquare,dineentop,randomwalk,phaserandomwalk,autox}.

In Fig.\ref{dist} we plot on the Argand plane the $\alm$ of the DILC
map for $\l \le 10$ (amplitudes $|\alm|$ in unit of $\mu$K.).
Because of the conjugate properties of the $\alm$ for a real sky
signal, we plot only $\alm$ modes with $m \ge 1$ and omit all $m=0$
modes. Note the apparent non-uniformity of the phases for $\l\le 5$.

\section{Testing the Random Phase Hypothesis}
We use Kuiper's statistic \citep{kuiper} (KS) to test on the random
phase hypothesis. The KS can be viewed as a variant of the standard
Kolmogorov--Smirnov test, designed to cope with circular data. The
standard Kolmogorov--Smirnov statistic is taken as the maximum
distance of the cumulative probability distribution against the
theoretical one: $D=\max_{-\infty < x < \infty}|S_N(x)-P(x)|$. For a
circular function, however, one needs to take into account the
maximum distance both above and below the theoretical probability
$P(x)$. Accordingly, we define: $D_{\pm}=\max_{-\infty < x <
\infty}\pm (S_N(x)-P(x))$. Then the test statistic is
\begin{equation}
V = D_{+}+D_{-}. \end{equation} In standard frequentist fashion, we
define the significance level $\alpha$ (or $p$-value or ``size'')
for our ``null'' hypothesis (of uniform randomness of the phase
angles) as the probability of the measured value of $V$ arising
under the null hypothesis. In this case, $\alpha$ can be calculated
from $\alpha=Q_{\rm Kuiper}(V [ \sqrt{N}+0.155+0.24/\sqrt{N}
  ] )$ and  $Q_{\rm Kuiper}(\lambda)=2\sum^{\infty}_{j=1}(4 j^2 \lambda^2-1)
e^{-2j^2\lambda^2}$, where $N$ is the number of data points.

We perform three statistical tests based on this general approach:
 on the {\it uniformity} of phases (i.e. consistency with a uniform distribution on the interval
 $[0,2\pi]$); on the {\it randomness} (i.e. independence) of phases by taking difference of phases with
fixed separation $(\dl, \dm)$;  and on the cross-correlation of each
$\l$ between DILC and the foregrounds by $\Delta
\phi^X_\lm=\phi^\dilc_\lm-\phi^{\rm FG}_\lm$. In each case the
resulting angles should be random: the difference between any two
random angles is itself a random angle.

\begin{inlinetable}
\begin{tabular}{||c|c|c|c||}
\hline  & $\phi_\lm$ of $\l \le 5$ & $\phi_\lm$ of $6\le \l \le 10$ & all $\phi_\lm$ of $\l \le 10$ \\
\hline ILC & 38.74\% & 90.91\% & 73.55\% \\
\hline DILC & 30.63\% & 98.09\% & 66.14\% \\ \hline
\end{tabular}
\caption{The significance level $\alpha$ in accord with the uniform distribution
hypothesis of the phases of the 1-year ILC and 3-year DILC maps.}
\end{inlinetable}

We first test the uniformity of the phases for two groups ($1 \le \l
\le 5$), and ($6 \le \l \le 10$) and for all phases claimed to
describe clean modes (i.e. $\l \le 10$). The phases of the $m=0$
modes are excluded in all our test. Table 1 shows that for the
3-year DILC, while the phases are consistent with a uniform
distribution ($\alpha\sim$ 0.98) for $6 \le \l \le 10$, and for $\l
\le 5$ ($\alpha \sim 0.31$). This is consistent with the behaviour
of the phases seen in Fig.\ref{dist}: the phases seem to be
concentrated in the  first two quadrants. This is however, only
significant at the 31\% level, which is barely $1\sigma$. Overall,
therefore, the phases of the 3-year DILC map for $\l \le 10$ are
consistent with uniformity ($\alpha=0.66$).

\begin{apjemufigure}
\centerline{\includegraphics[width=0.9\linewidth]{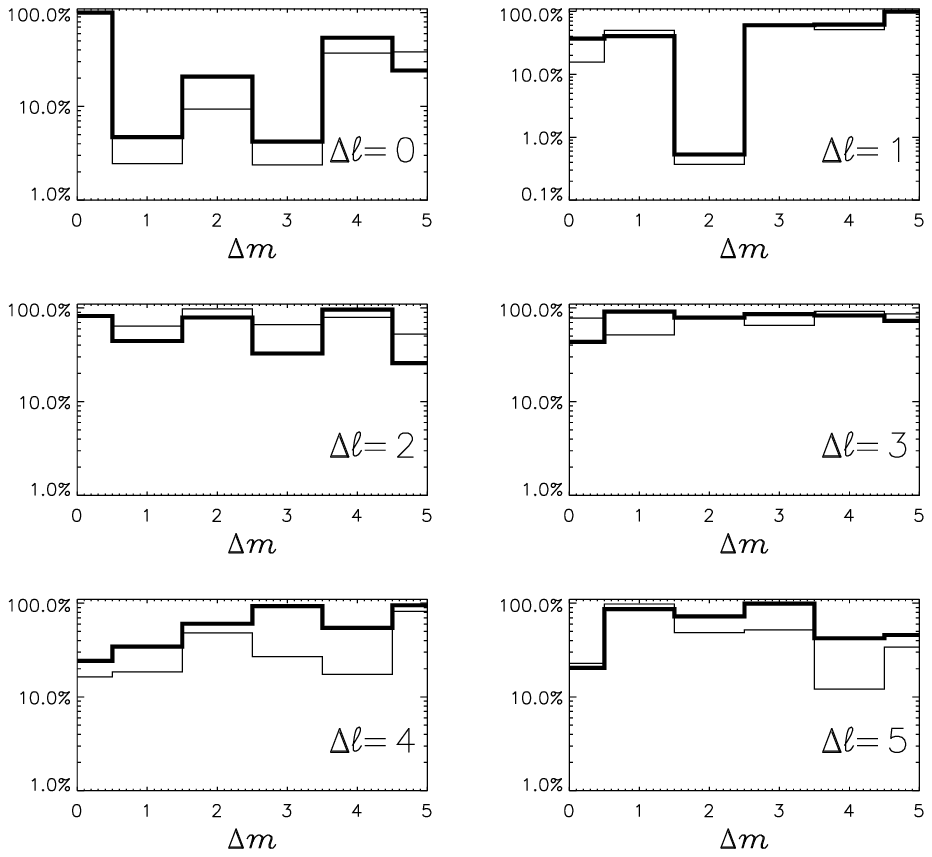}}
\caption{Significance levels of randomness with separation
$(\dl,\dm$). The thick and thin lines are from  the 3-year DILC and
1-year ILC maps, respectively.} \label{auto}
\end{apjemufigure}

Now we check the randomness of the phases is tested by the defining
a set of difference $\Delta \phi^\dilc(\dl,\dm)=\phi^\dilc_{\l+
\dl,m+\dm}-\phi^\dilc_\lm$. In Fig.\ref{auto} we show the
significance levels for randomness between phases with different $l$
and $m$. There are three separations that show significant
departures from uniformity ($\alpha=0.047$, $0.0421$ and $0.0053$
for $(\dl,\dm)=(0,1)$, $(0,3)$ and $(1,2)$, respectively. This
corresponds to significant coupling of the phases. It is complicated
to understand coupling across both $\l$ and $m$ for
$(\dl,\dm)=(1,2)$ because of the lack of rotational invariance
described earlier. Nevertheless, because the first two examples
involve coupling between azimuthal numbers $m$ within each $\l$, we
plot in Fig.\ref{phsdif} the sequences of phase difference $\Delta
\phi^\dilc(0,1)$ and $\Delta \phi^\dilc(0,3)$ on a unit circle (or
$\exp(i\Delta \phi^\dilc)$ on an Argand plane). For both
distributions one can see the deficits around $\Delta \phi=0$ that
cause the significance levels for the cases to appear below 5\%. We
also include $(0,2)$ in Fig.\ref{phsdif} for comparison, which has
the apparent tendency of phase differences to avoid $\pi$, though
this is not significant at the $5$\% level.

We would like to emphasize that the statistics derived from these 35 
separations shown in Fig.2 should not be treated as a statistical 
ensemble. For example, primordial cosmological magnetic field induces 
and supports vorticity or Alfv\'en waves, which induce in CMB anisotropies 
with correlation between  $a_{\l+1,m}$ and $a_{\l-1,m}$, i.e. $\Delta \l=2$ 
\citep{dky,chen}. Therefore, examining $\Delta 
\l=2$ correlation alone is qualified as an independent method. Another 
example is that symmetric signals defined on the Galactic coordinate 
system with respect to the Meridian ($\varphi=0$) on a sphere induce 
correlations between $\Delta \l=4$ \citep{4n,2n}. Therefore, each of the $(\Delta \l, \Delta m)$ should be treated 
as a separate non-Gaussianity test like bispectrum and trispectrum\ldots etc..

\section{Cross-correlation between the DILC and foreground maps}
Since the DILC map is obtained from an internal combination of the
frequency maps, some foreground residues might be left unsubtracted.
Based on the assumption that CMB signal should not correlate with
the foregrounds, and that the characteristic of phases reflect the
morphology of the CMB anisotropy pattern, we also test the
cross-correlation of phases for the DILC and the \wmap 3-year
foreground maps at K, Ka, Q, V and W channels. The foreground maps
we test are the sum of the synchrotron, free-free and dust
templates. We take the phase difference $\Delta
\phi^X_\lm=\phi^\dilc_\lm-\phi^{\rm FG}_\lm$ for each $\l$, assuming
such $\Delta \phi_\lm$ at each $\l$ should be uniformly distributed.
In Fig.\ref{xcorr} one can see that for $\l=3$ and 6  the DILC
phases have correlation with the foregrounds with significance
around 10\%.  One particular point about the quadrupole is easily
seen from Fig.\ref{dist}: two of the three quadrupole phases are
near 0 and $\pi$. Note also that, for $\l=6$, in Fig.\ref{phsdif}
(shown in red diamond sign), the phase differences for $\dm=1$ are
strongly clustered.

\begin{apjemufigure}
\centerline{\includegraphics[width=1.0\linewidth]{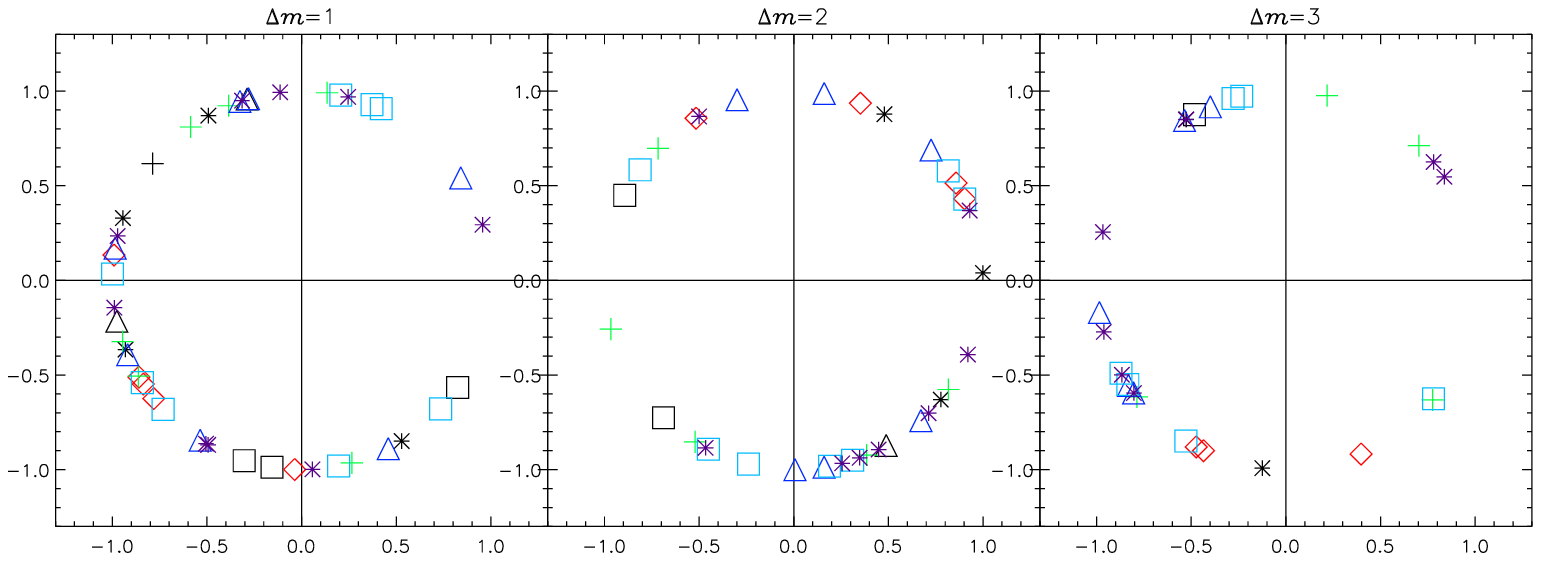}}
\caption{Phase difference for $\dm=1$, 2 and 3 for all
$\phi^\dilc_\lm$, $\l\le10$, $1 \le m \le \l$ plotted on a unit
circle. The significance levels of departures from uniformity are
4.70\%, 20.85\% and 4.21\%, respectively. The signs  $\diamond$,
$+$, $\triangle$, $\Box$ and $\ast$ in black represent phase
difference within $\l=1-5$, respectively, and in color within
$\l=6-10$, respectively.} \label{phsdif}
\end{apjemufigure}

\section{Extrema Statistics}
The $\dm$ phase coupling in each $\l$ leads to a departure from
Gaussianity in the resulting signal. Following \citet{autox} we
represent the effect of phase coupling by assembling the $\alm$ (now
only single variable $m$) with an inverse Fourier transform:
\begin{equation}
\Delta T_\l(\varphi)=\sum_m \alm^\dilc \exp(i m \varphi),
\end{equation}
where the negative $m$ in the sum are replaced with $\alm^\ast$ to
ensure that $\Delta T_\l$ is real. The
morphology of the signal obtained by this method is similar to the
signal $\Delta T(\theta,\varphi)=\sum_{m=-\l}^\l \alm \ylm(\theta,\varphi) $ obtained for each $\l$ by summing over all $\theta$ onto $\varphi$ axis on the map. One should note the following subtleties of such comparison. The map can be written
\begin{eqnarray}
\Delta T_\l&=&\sqrt{\frac{2\l+1}{4\pi}} \left\{ a_{\l 0} P_{\l 0}(\cos\theta) + 2\sum_{m=1}^\l    \left[ (-1)^m \right.\right. \nonumber \\
&& \left.\left. \sqrt{\frac{(\l-m)!}{(\l+m)!}} |\alm|\cos(m\varphi +\phi_\lm) P_\lm(\cos \theta)\right]\right\}.
\end{eqnarray}
The integration of the associated Legendre function is zero for odd $\l+m$. When one compares the 1D curve with the composite map for odd $\l$, only the odd $m$ contribute to the integration while the $(-1)^m$ reverses the signal as seen from simple inverse Fourier transform $\sum |\alm| \cos(m\varphi +\phi_\lm)$.   

Such 1D construction obviously loses information in one of the available dimensions, but
despite the fact that the $\alm$ are spherical harmonic
coefficients, the statistics registered in the one-dimensional
complex $\alm$ should still manifest themselves in the 1D $\Delta T$
curves we create. In Fig.\ref{1d} we plot $\Delta T(\varphi)$
summing from the DILC $\alm$ (top), and from the whitened DILC:
$\alm/|\alm|$ (bottom). If the signal is Gaussian, the locations of
the highest and lowest peaks should randomly distributed in
$\varphi$ between $-180^\circ$ and $180^\circ$. We plot in
Fig.\ref{peakoncircle} the distribution of these extrema locations
in a unit circle.

\begin{apjemufigure}
\centerline{\includegraphics[width=0.9\linewidth]{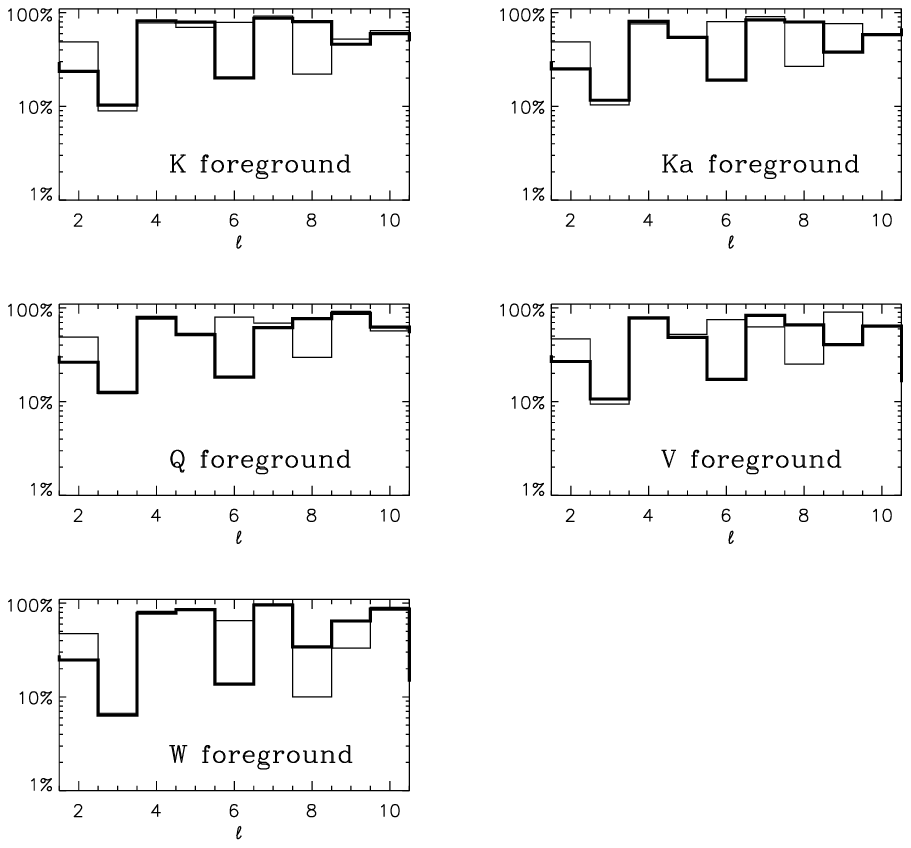}}
\caption{Significance levels for the cross-correlation of phases at
each $\l$ between the DILC (thick lines) and the \wmap foreground
maps, from top to bottom, K, Ka, Q, V and W channels, respectively.
For comparison, the thin lines are those between 1-year ILC and
1-year foreground maps.} \label{xcorr}
\end{apjemufigure}

In Table 2 we list the significance of the distribution of the
extrema locations. Not only do they show low significance levels of random distribution in $\varphi$, the peak locations also cluster in $\varphi=180^\circ$. In Fig.\ref{xcorr} the $\l=3$ and 6 show
significant cross correlation with the foregrounds. We therefore
test the peak distribution by excluding the four extrema belonging
to these two modes. Although $p$-values increase slightly, the
results are still significant at a level below 5\%.

\begin{inlinetable}
\begin{tabular}{|c|c|c|}
\hline                   & 3-y DILC    &  Whitened 3-y DILC \\
\hline all peaks $\l \le 10$        & 0.450\% &  0.552  \%   \\
\hline excluding peaks of $\l=3, 6$ & 2.759\% &  4.218 \% \\  \hline
\end{tabular}
\caption{Significance levels of the uniformity of the distribution
of the extrema locations in $T_\l$.}
\end{inlinetable}

\begin{apjemufigure}
\centerline{\includegraphics[width=1.\linewidth]{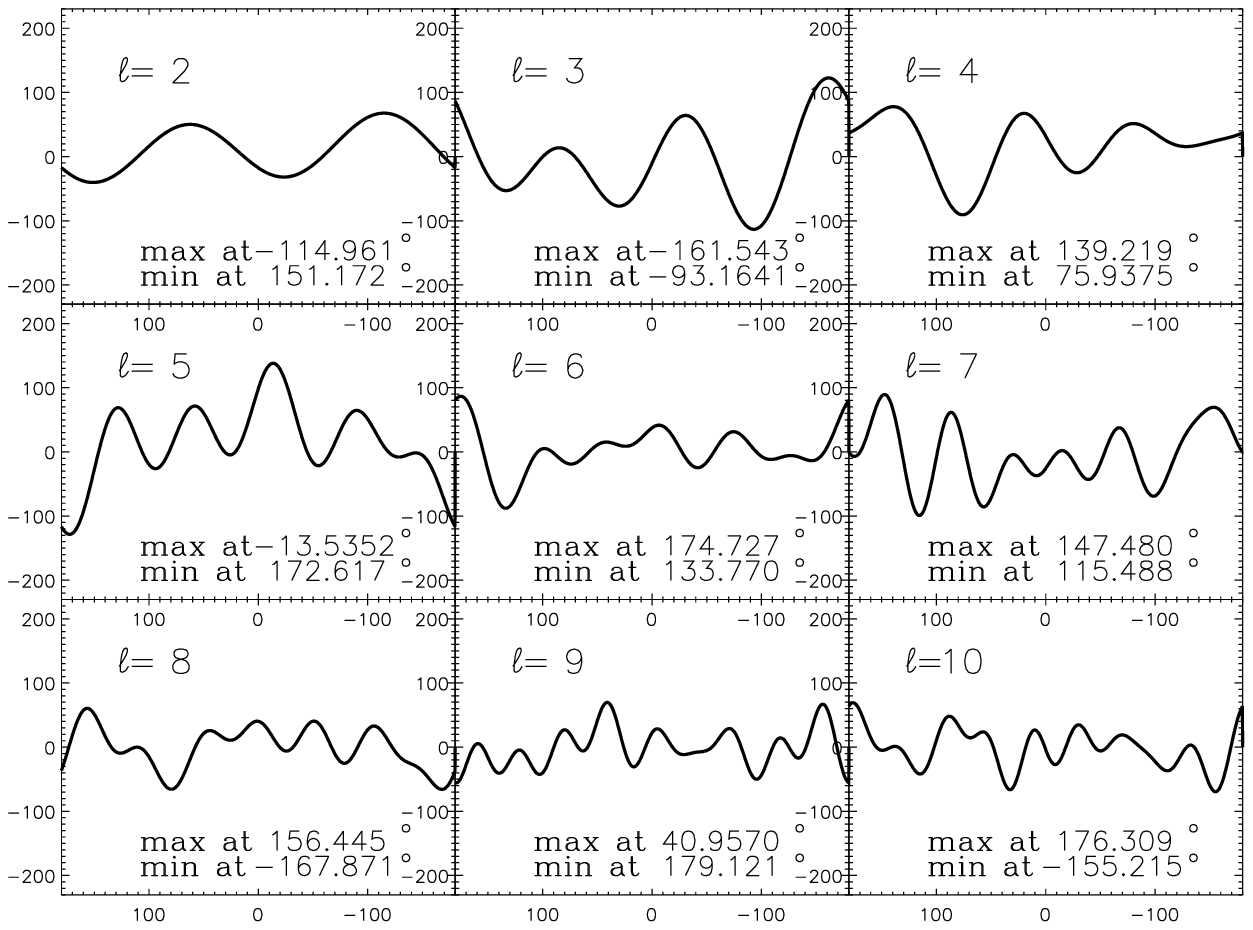}}
\centerline{\includegraphics[width=1.\linewidth]{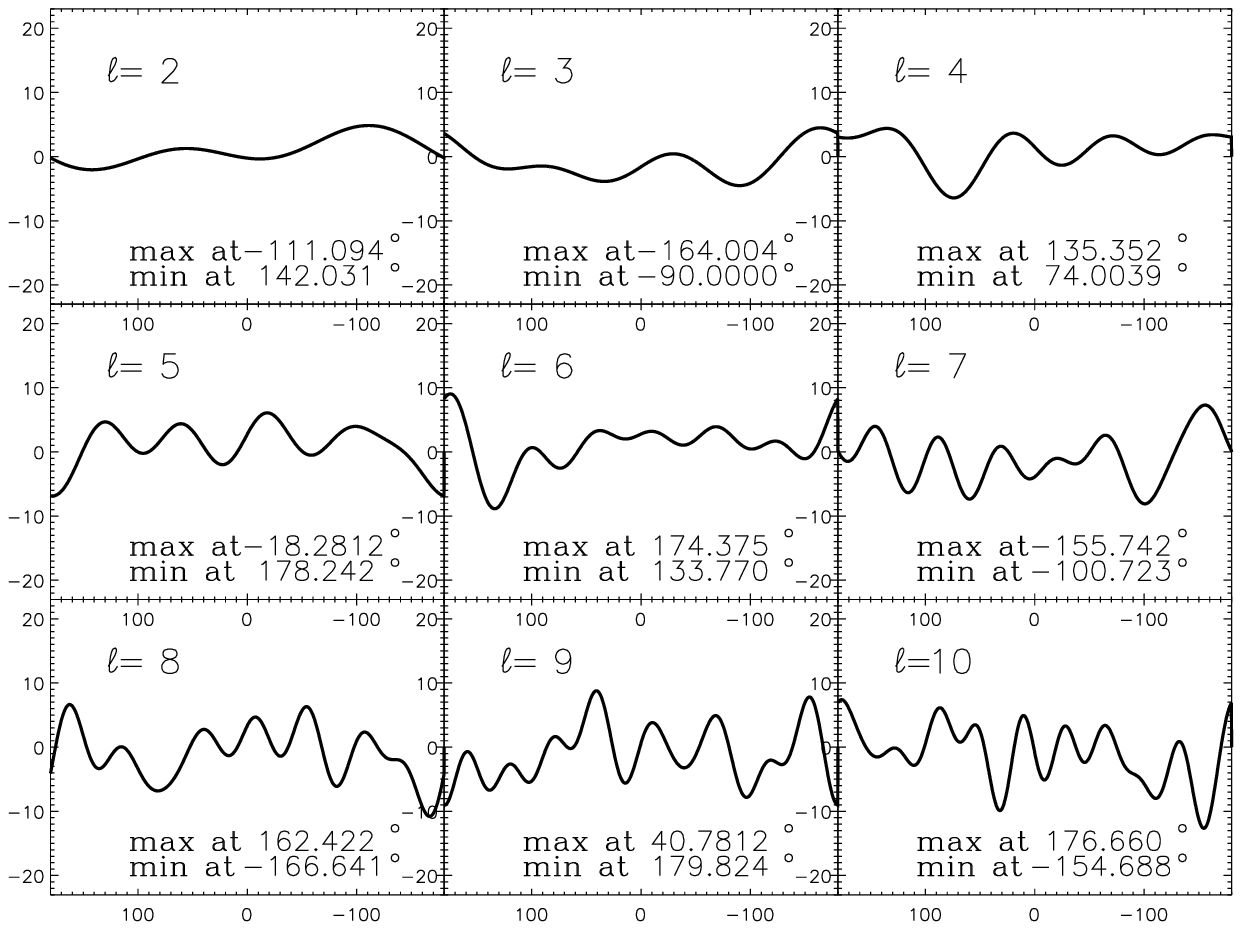}}
\caption{The $\Delta T_\l(\varphi)$ distribution for each $\l$ from
the DILC map by assembling $\sum_m \alm^\dilc \exp(i m \varphi)$
(top panel, in which the unit of $y$ axis is $\mu$K) and by
assembling only the phase part: $\sum_m \exp(i \phi^\dilc_{\lm})
\exp(i m \varphi)$ (bottom). In order to match the convention of
Galactic longitude coordinates (for comparison with, e.g. Fig.14 in
\citet{wmap3ytem}), the $\varphi$ axis is plotted reversely. Note that on comparison, due to the properties of spherical harmonics, peaks on the maps for odd $\l$ correspond to troughs on the 1D curves. In each
figure we indicate the locations of the extrema.} \label{1d}
\end{apjemufigure}

\section{Conclusion}
In this letter we test the Gaussian random hypothesis of the CMB
temperature anisotropies. The behaviour of the 3-year DILC does not
differ strongly from that of the 1-year ILC version: all the famous
peculiarities still exist, as mentioned in \citet{wmap3ycos}. We
have found in this paper that this also extends to the behaviour of
the spherical harmonic phases. In particular, we find that phase
differences (which should be uniformly distributed), tend to avoid
the region of the complex plane close to $\Delta \phi=0$. We also find
that the phases for $\l=3$ and $\l=6$ are significantly correlated
with those of the foreground maps. We also test the real space
alignment of the resulting features using the temperature extrema
resulting from a Fourier summation for each $\l$. The resulting
peaks are indeed concentrated at opposite to the center of our Galaxy, i.e. $l=180^\circ$ of Galactic
coordinates, with respect to which the phases are themselves
defined.

On the basis of these results we reject the null hypothesis that the
modes with $l\leq 10$ are a realization of a statistically
homogeneous Gaussian random field at a significance level better
than 5\%. We also infer that the origin of the observed departures
is consistent with being some form of Galactic foreground.

Of course it is possible the apparent alignment between CMB
temperature pattern and galactic foreground morphology is simply
fortuitous. It could be that large-scale anisotropies in both line
up accidentally. If this is the case then we just happen to live at
a place in the Universe where our past light cone presents us with
this coincidence and we will just have to cope with it; all future
diagnostics of foreground contamination will have to incorporate
this alignment as conditioning information. We believe however that
it is important to take such coincidences very seriously until we
know for certain that is all they are.

\begin{apjemufigure}
\centerline{\includegraphics[width=1.\linewidth]{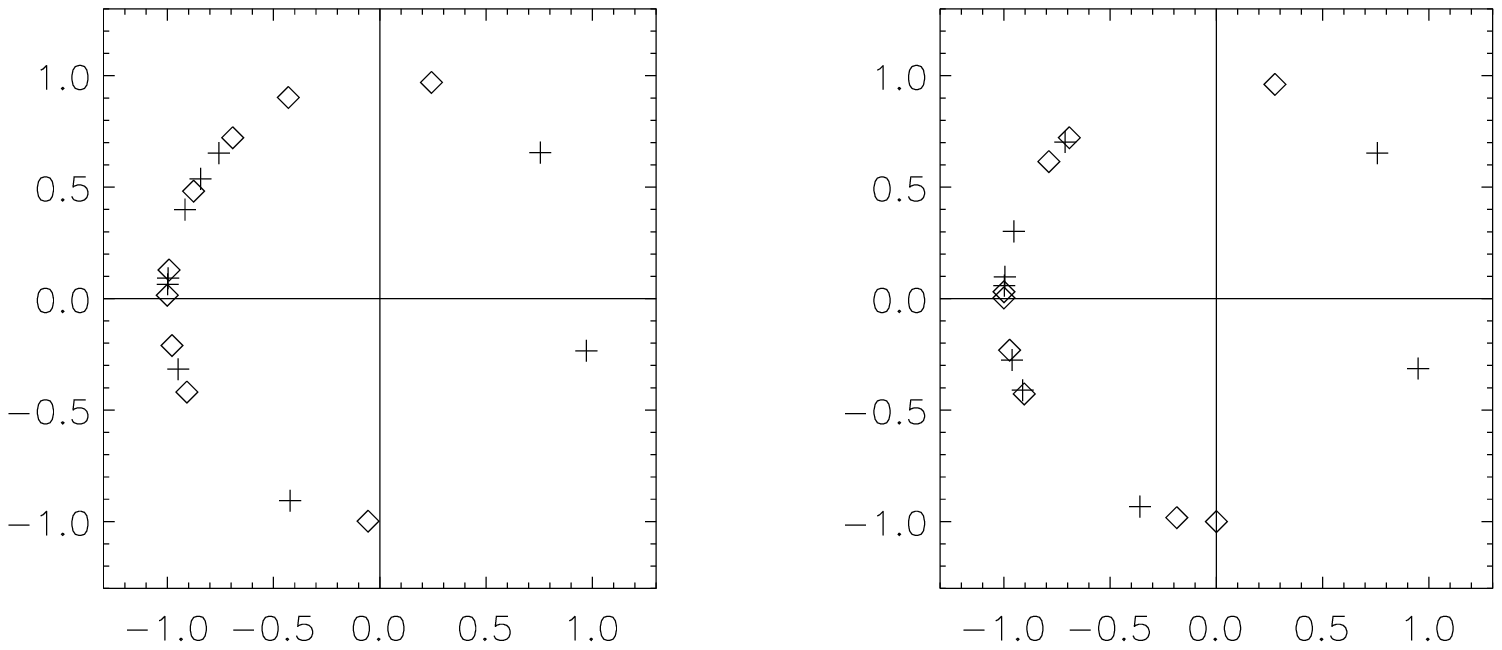}}
\caption{The distribution of the extrema locations of $\Delta T_\l(\varphi)$ as shown in Fig.\ref{1d}.
The angle of the unit circle denotes the location $\varphi$.
The left panel is for the DILC $\Delta T_\l(\varphi)$, and the right for
the whitened DILC $\Delta T_\l(\varphi)$. The $+$ sign denotes maxima and $\diamond$ sign minima.}
\label{peakoncircle}
\end{apjemufigure}

\section*{Acknowledgments}
We acknowledge the use of the NASA Legacy Archive for extracting the \wmap data.
We also acknowledge the use of \healpix \footnote{\tt http://www.eso.org/science/healpix/} package
\citep{healpix} to produce $\alm$ from the \wmap data and the use of \glesp \footnote{\tt http://www.glesp.nbi.dk} package \citep{glesp}.

\end{document}